\documentclass{emulateapj} 

\newcommand{\emm}[1]{\ensuremath{#1}}   
\newcommand{\emr}[1]{\emm{\mathrm{#1}}} 

\slugcomment{Submitted 2007 October 5; accepted 2008 February}
\shorttitle{Star Formation Near Photodissociation Regions}
\shortauthors{Goicoechea et al.}

\begin{document}

\title{Star Formation Near Photodissociation Regions:\\ 
Detection of a Peculiar Protostar Near Ced~201\altaffilmark{1,2,3}}


\author{Javier R. Goicoechea$^4$, Olivier Bern\'e$^5$, Maryvonne Gerin$^4$, 
Christine Joblin$^5$, David Teyssier$^6$}

\affil{$^4$LERMA-LRA, UMR 8112, CNRS, Observatoire de Paris et 
\'Ecole Normale Sup\'erieure, 24 rue Lhomond, 75231 Paris Cedex 05, 
France.\\}

\affil{$^5$Centre d'Etude Spatiale des Rayonnements, CNRS et Universit\'e Paul 
Sabatier Toulouse~3, Observatoire Midi-Pyr\'en\'ees, France.\\}

\affil{$^6$European Space Astronomy Centre, ESAC, Urb. Villafranca del 
Castillo, P.O. Box 50727, Madrid 28080, Spain.}

\altaffiltext{1}{Based on observations obtained with the CSO telescope. 
The CSO telescope is operated by the California Institute of Technology under funding
from the NSF, Grant No. AST-0540882.}
\altaffiltext{2}{Observations also obtained with IRAM telescopes,
supported by INSU/CNRS (France), MPG (Germany), and IGN (Spain).} 

\altaffiltext{3}{This work is based on observations made with the Spitzer Space Telescope, 
which is operated by the Jet Propulsion Laboratory, 
California Institute of Technology under a contract with NASA.}

\begin{abstract}

We present the detection and characterization of a peculiar low--mass protostar (IRAS~22129+7000)
located $\sim$0.4\,pc from Cederblad~201 Photodissociation
Region (PDR) and $\sim$0.2\,pc from the HH450~jet.
The cold circumstellar envelope surrounding the object has been 
mapped through its 1.2~mm dust continuum emission with \textit{IRAM--30m}/MAMBO. 
The deeply embedded protostar is clearly detected 
with \textit{Spitzer}/MIPS (70~$\mu$m), IRS (20--35~$\mu$m) and IRAC (4.5, 5.8, and 8~$\mu$m) 
but also in the $K_s$ band (2.15~$\mu$m). 
Given the large \textit{near-- and mid--IR excess} in its spectral energy distribution,
but large submillimeter-to-bolometric luminosity ratio ($\simeq$2\%),
IRAS~22129+7000 must be a transition Class~0/I source and/or a multiple stellar system.
Targeted observations of several molecular lines from 
CO, $^{13}$CO, C$^{18}$O, HCO$^+$ and DCO$^{+}$ have been obtained.
The presence of a collimated molecular outflow mapped with the \textit{CSO}
telescope in the CO $J$=3--2 line
suggests that the protostar/disk system is still accreting material from its natal envelope.
Indeed, optically thick line profiles from high  density tracers such as HCO$^+$ $J$=1--0
show a red--shifted--absorption asymmetry  reminiscent of inward motions.
We construct a preliminary physical model of the circumstellar envelope 
(including radial density and temperature gradients, velocity field and turbulence)
that reproduces the observed line profiles and estimates the
ionization fraction. The presence of both mechanical and (non-ionizing) FUV--radiative input makes
the region an interesting case to study triggered star formation.%

\end{abstract}


\section{Introduction}

Low--mass  stars (sunlike; M$_{\star}$\,$<$8\,${M_\odot}$) form from the gravitational collapse of
dense molecular cores within giant molecular clouds. 
During the earliest stages of evolution, embryonic protostars are  
deeply embedded in cold and dusty envelopes of infalling material.
In recent years, large millimeter (mm) and submm continuum surveys
have been carried out to detect the cold dust emission associated
with these early birth-sites of star formation \cite{And93,Joh00}.
During collapse, conservation of angular momentum 
combined with infall along the magnetic field lines
leads to the formation of a rotating protoplanetary disk that drives the 
accretion  process. 
At the same time, both mass and angular momentum are removed from the system by the onset
of jets, collimated flows and the magnetic braking action \cite{Bon96,Cab07}. 
The resulting molecular outflow starts to erode and sweep up part of the natal envelope,
contributing for the clearing of the circumstellar material
and the termination of the infall phase \cite{Arc06}. The so--called
Class~0 sources are deeply embedded protostars at their  main
accretion phase and represent the earliest type of young stellar objects (YSOs).
Observationally they should show: $(i)$ Evidence of a central YSO
(as revealed by the presence of a collimated outflow or an internal heating source);
$(ii)$ Centrally peaked but extended massive envelope (as traced
by mm and submm dust continuum emission); $(iii)$ High submm ($\lambda$\,$>$350\,$\mu$m) to
bolometric luminosity ratio (L$_{smm}$/L$_{bol}$\,$>$0.5$\%$), 
and a spectral energy distribution (SED) similar to a single graybody at 15-30\,K \cite{And00}. 
On the other hand,
Class~I sources correspond to the late accretion phase (roughly half of the
initial envelope mass is accreted in the Class~0 stage), they don't
satisfy $(iii)$ and  can easily be detected at near--IR wavelengths (Lada 1999).
 
The detection of YSOs that do not  exactly match  
the above canonical categories can provide the missing pieces of our current understanding of 
early disk evolution and early dissipation of the circumstellar envelope.
These sources can thus potentially help to address, for example, how and when the velocity field around
YSOs changes from infall to rotation dominated, which fraction of the initial mass envelope is 
dissipated and not accreted onto the star/disk, as well as the timescales for such dissipation
\cite{Bri07,Jor07}

\begin{figure*}[ht]
\centering
\includegraphics[angle=0,width=14.5cm]{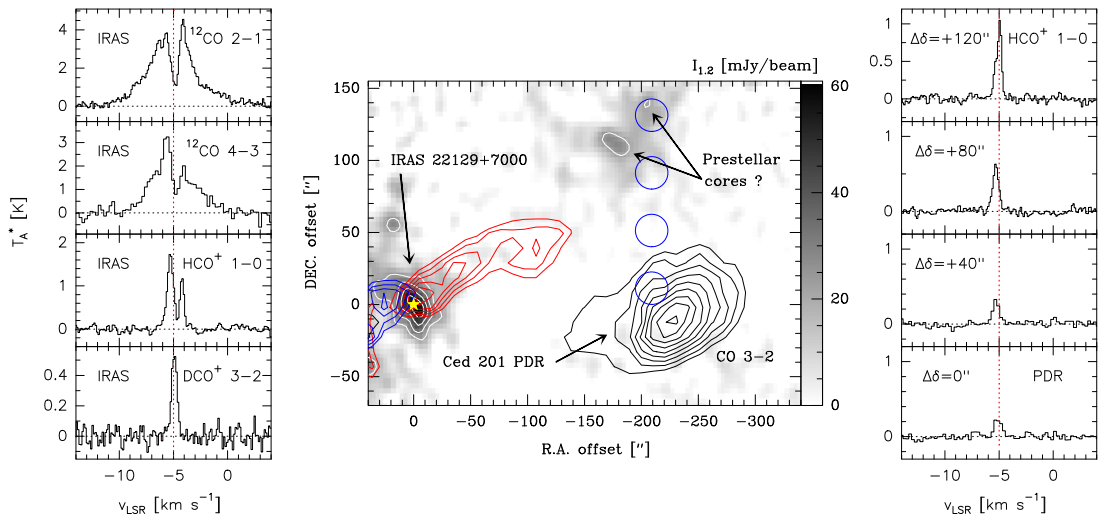}
\caption{\textit{Middle panel:} IRAM--30m/MAMBO 1.2~mm-continuum emission of the region.
The image is centered at IRAS~22129+7000. 
White contours are displayed from 22 (2$\sigma$)
to 55~mJy/15$''$-beam in steps of 11~mJy/15$''$-beam (1$\sigma$).
The CO $J$=3-2  integrated intensity (from 5 to 9~K~km~s$^{-1}$ in steps of 1~K~km~s$^{-1}$) 
from the outflow
is shown in  red (from $v$=-4 to -2~km~s$^{-1}$) and  blue (from $v$=-6 to -8~km~s$^{-1}$). 
The integrated CO $J$=3-2 line intensity 
from Ced~201 PDR ($v$=-5.5 to -4.5~km~s$^{-1}$) is shown in black contours 
(from 6 to 20~K~km~s$^{-1}$ in steps of 1~K~km~s$^{-1}$).
\textit{Left panels:} $^{12}$CO $J$=2-1 and 4-3, HCO$^+$ $J$=1-0 and 
DCO$^+$ $J$=3-2 lines towards IRAS~22129+7000 position. 
\textit{Right panels:} Declination raster in the HCO$^+$ $J$=1-0 line. 
The vertical dashed line marks the rest velocity of the cloud ($-$5\,km\,s$^{-1}$).}
\label{fig:mambo}
\end{figure*}

In this paper we present the detection of a low--mass protostar in the Bok globule B175 (L1219)
within the Cepheus Flare, a nearby ($\sim$400~pc) star forming region
$\sim$85\,pc above the galactic plane \cite{Kun98,Nik04}.
This cloud hosts the enigmatic Cederblad~201  
reflection nebula \cite{Wol08, Ces00}, probably the result
of a chance encounter  with a B9.5V runaway star (BD+69$^\circ $1231, T$_{eff}$\,$\simeq$10,000~K),
which is moving through the cloud at $\sim$12\,km\,s$^{-1}$ \cite{Wit87}.
The star is thus not a only a source of dissociating photons
(the FUV radiation field is $\sim$200 times the mean
interstellar radiation field) but also  a source of 
shocks and turbulence in the region. This interaction enables us to probe matter 
in a different situation than in classical PDRs
in which the distance from the exciting star to the illuminated  regions is large
and/or in which the star was formed locally. 
In addition, the extended cloud hosts a  Herbig--Haro jet (HH450), and several arsec-scale 
filaments of the  supernova remnant (G110.3+11.3) that are  rapidly approaching to 
the region \cite{Bal01}.
The presence of both mechanical (from shocks and outflows) and radiative input
(from FUV photons) makes the region an interesting
dynamics laboratory to study how star formation is proceeding through the cloud.

\begin{deluxetable}{cccc}
  \tablecaption{IRAS~22129+7000 Photometric data\label{tab}}
    \tablecolumns{4}
    \tablehead{
      \colhead{$\lambda$} &
      \colhead{Flux} &
      \colhead{Telescope} &
      \colhead{Aperture}\\
      \colhead{($\mu$m)} &
      \colhead{(Jy)} &
      \colhead{} &
      \colhead{}\\}
\startdata
   1.25 & $^{\dagger}$3.20$\times10^{-4}$   & 2MASS & 3$''$\\
   1.65 & $^{\dagger}$3.60$\times10^{-4}$   & 2MASS & 3$''$\\
   2.15 & 3.00$\times10^{-3}$   & 2MASS & 3$''$ \\
   4.5  & 0.17   &  Spitzer/IRAC & 12.2$''$\\
   5.8  & 0.38   &  Spitzer/IRAC & 12.2$''$\\
   8.0  & 0.31   & Spitzer/IRAC & 12.2$''$\\
   12   & $^{\dagger}$0.39   & IRAS & 5$'$\\
   25   & 1.13  & IRAS & 5$'$\\
   60   &  3.46 & IRAS & 5$'$\\
   70   &  6.10 & Spitzer/MIPS & 1$'$ \\
   100  & 16.00 & IRAS &5$'$\\
   200  & 22.00 & KAO & $\sim$1$'$ \\
   1200 & 0.25  & IRAM-30m/MAMBO & 31$''$$\times$22$''$$^*$\\
\enddata
\tablecomments{$^{\dagger}$Upper
limit. $^*$Source FWHM size given.}
\label{table: fluxes}
\end{deluxetable}

\section{Observations}

A 1.2\,mm continuum emission map of the region  was obtained at the 
\textit{IRAM-30m} (Pico Veleta, Spain) in March 2005  using the 117--channel 
MAMBO bolometer array.
The angular resolution is $\sim$11$''$. A fast--mapping mode 
was used to map the region \cite{Tey99}. The total integration time was 1\,hour,
achieving a rms noise of 7\,mJy/11$''$-beam.
Sky noise substraction and  data analysis were carried out with the \texttt{MOPSIC}
software \cite{Zyl98}.  
A 31$''\times$22$''$ full-width at half-maximum (FWHM)  
condensation with  a peak intensity of 60$\pm$7\,mJy/11$''$-beam 
is detected $\sim$225$''$ East of Ced~201 PDR.
The integrated flux  is 249$\pm$50\,mJy, which reflects the estimated
absolute calibration uncertainty. 
The resulting 1.2\,mm emission map is shown in Figure~\ref{fig:mambo} 
(smoothed to a beam resolution of 15$''$).

\begin{figure*} [ht] 
\centering
\includegraphics[angle=0,width=13.9cm]{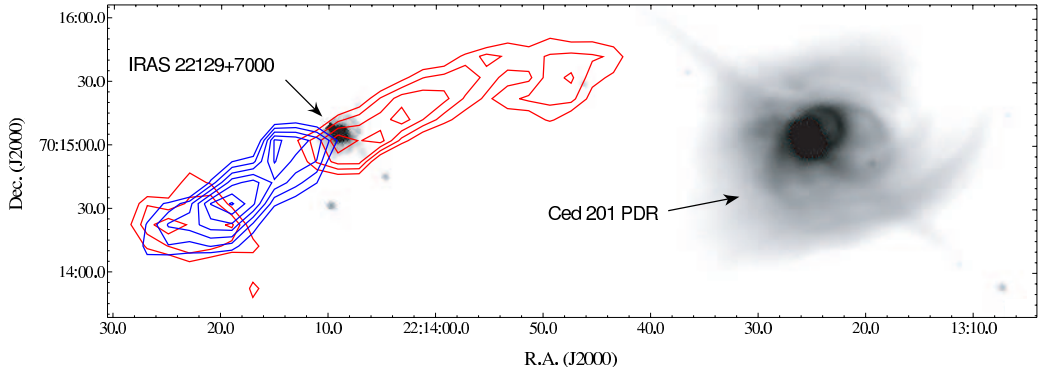}
\caption{\textit{Spitzer}/IRAC 8\,$\mu$m image of Ced201 region including
the detection of IRAS~22129+7000 protostar.
The bright reflexion nebula (the PDR) is clearly seen at the west of the region.
A well collimated outflow (as traced by CO $J$=3--2
high velocity line wing emission) emerges from  the protostar position.
Contours show the integrated intensity (from 5 to 9 K~km~s$^{-1}$ in steps of
1 K~km~s$^{-1}$) in the red (from $v$=-4 to -2~km~s$^{-1}$) and  blue (from $v$=-6 to -8~km~s$^{-1}$) 
line wings respectively.}
\label{fig:IRAC}
\end{figure*}

The HCO$^+$ $J$=1--0 line was observed also at
the --30m telescope during September 2005 using the B100 single sideband
receiver with a channel resolution of 40\,kHz. 
The main beam efficiency at 89\,GHz is 0.82 and the angular resolution
is 28$''$. The line was observed towards the dust condensation and in a
declination raster crossing Ced~201 PDR (see Fig.~\ref{fig:mambo}).
Additional observations  were performed at the CSO  telescope (Mauna Kea, Hawaii) during
July 2006 and June 2007. 
The telescope is equipped with SIS receivers operated in double sideband mode. 
The  CO $J$=3--2 map (Figs.~\ref{fig:mambo} and ~\ref{fig:IRAC}) was made 
``on-the-fly".  The DCO$^+$ $J$=3--2 line was mapped in a $\sim$40$''$\,$\times$30$''$
region around the dust condensation, whereas pointed observations 
in the CO $J$=2--1 and 4--3 lines were taken 
towards the dust emission  peak (Fig.~\ref{fig:mambo}). 
The spectra were analyzed with a 1024 channel
acousto-optic spectrometer with a total bandwidth of 50~MHz and a 
resolution of $\sim$100\,kHz.
The main beam efficiencies of the telescope were 0.70, 0.75 and 0.53
at 230, 345, and 461\,GHz respectively. The angular resolution 
is $\sim$30$''$ at CO $J$=2--1, $\sim$20$''$ at CO $J$=3--2 and
$\sim$15$''$ at CO $J$=4--3.
Data was processed with \texttt{GILDAS}.

The Ced~201 region was also observed with \textit{Spitzer} as part of our 
SPECPDR program\footnote{http://www.cesr.fr/$\sim$joblin/SPECPDR\_public/Home}
\citep{Job05} with the Infrared Array Camera (IRAC) at 4.5, 5.8, and 8~$\mu$m
and with the  Infrared Spectrograph (IRS).
IRS data were obtained in the spectral mapping mode with the LL1 module 
(spectral range: 20 to 35\,$\mu$m).
The spectral cube of the source was assembled using the CUBISM software \citep{Smi07} from the 
Basic Calibrated Data (BCD) files.  
The 20 to 35\,$\mu$m data was obtained by integrating the cube spatially (see Bern\'e et al. 2007).
The 70~$\mu$m Multiband Imaging Photometer for Spitzer
(MIPS) data that are part of the PID 30790 program (Fazio et al.)
were  retrieved from the \emph{Spitzer} archive.
MIPS and IRAC post-BCD calibrated images were used directly
without additional processing.

\section{Results: Detection of a Young Protostar}

Figure \ref{fig:IRAC} shows the  detection of a compact  
continuum source at $\sim$8\,$\mu$m (warm dust) located $\sim$0.4\,pc East of  Ced~201 PDR \cite{Ber07}.
The compact object is also detected at 4.5, 5.8, 20--35 and  70~$\mu$m with \emph{Spitzer}.
The  near--IR source position agrees with the continuum  peak of the 
1.2~mm extended emission (cold dust) detected in a larger field-of-view (Fig.~\ref{fig:mambo}).
The best coordinates inferred from IRAC images are $\alpha_{2000} 
= 22^h14^m08.3^s$, $\delta_{2000} = +70{^\circ} 15' 06.7''$,
which coincides with the IRAS~22129+7000 source, clearly a YSO.
A well collimated molecular outflow  emanating
from the object is seen along a SE-NW axis in the CO $J$=3--2 line
(Figs.~\ref{fig:IRAC} and \ref{fig:mambo}). This outflow was previously detected in CO $J$=1--0 at
lower angular resolution by Nikoli\'c \& Kun (2004), who first suggested that  
IRAS~22129+7000 could be the driving source.
We now detect the near-- and mid--IR scattered light from the embedded protostar and the
mm thermal emission from the surrounding dusty cocoon.
In the following we examine the nature of this source
and whether or not the presence of a PDR plays a role in the star formation
process in the region.

\subsection{Peculiar SED Characteristics}

\begin{figure}[ht]
\centering
\includegraphics[angle=-90,width=8.5cm]{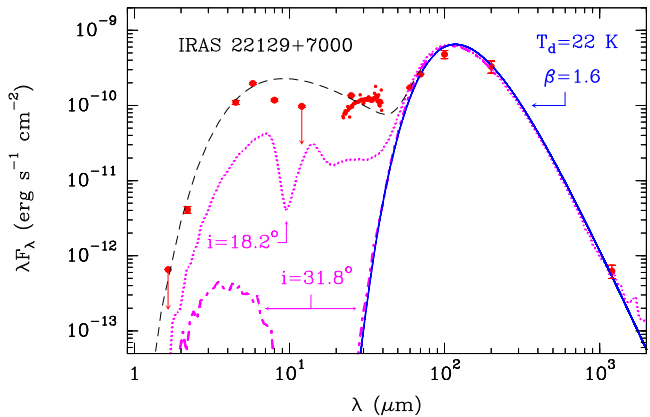}
\caption{SED of IRAS~22129+7000 combining data 
in the H and K$_s$ bands (2MASS; Skrutskie et al.~2006); 4.5, 5.8 and 8\,$\mu$m (Spitzer/IRAC);
20-35\,$\mu$m (Spitzer/IRS, smaller points), 12, 25, 60 and 100\,$\mu$m (IRAS), 70\,$\mu$m 
(Spitzer/MIPS), 200\,$\mu$m (KAO: Casey 1991) and 1200\,$\mu$m  (MAMBO). 
The Spitzer/IRS fluxes towards the source agree with those measured by IRAS at 25$\mu$m
probing the compact nature of the mid--IR emission.
The dashed curve is a simple fit using two graybodies
with different temperatures. The continuous blue curve represents the circumstellar
envelope with T$_d$=22\,K and a spectral index of 1.6.
The magenta pointed-- and pointed-dashed curves show the best 2D radiative transfer
models obtained with the Robitaille et al. (2007) fitter for two different
inclinations ($i=0$ corresponds to ``pole-on'' and $i=90$ to ``edge-on'' configurations).}
\label{fig:sed}
\end{figure}

In order to investigate the emitted power in the different wavelength domains,
the SED of IRAS~22129+7000  
has been built with our MAMBO and \textit{Spitzer} observations and complemented with 
previous 2MASS (2.15\,$\mu$m), IRAS (12, 25, 60 and 100\,$\mu$m)  and KAO 
(200\,$\mu$m; Casey 1991) detections (Table~\ref{table: fluxes}). 
We fitted the SED with two graybodies, 
$B_{\lambda}(T)\,(1-e^{-\tau_{\lambda}})\,\Omega$, where $\Omega$ is the solid angle subtended
by the emitting region. In the far-IR and mm domain 
$\tau_{\lambda}\propto\kappa_{1200}\left(1200/\lambda\right)^{\beta}$
and T can be associated  with the dust opacity and color temperature ($\simeq$T$_d$) 
of the extended envelope. Here we take a dust opacity 
(per gas+dust mass column density) of
$\kappa_{1200}$=0.01\,cm$^2$\,g$^{-1}$ at 1200\,$\mu$m, the usual value 
adopted for the  dust in protostellar envelopes \cite{And00}.
Because of the implicitly assumed  grain growth and ice mantle formation \cite{Oss94},
this  value is a factor $\sim$5 larger than $\kappa_{1200}$ for standard interstellar grains.
$\Omega$ is fixed as the FWHM size of the envelope inferred from the 1.2\,mm continuum map,
while the column density of material is varied between A$_V$\,$\simeq$5 and 30.
Satisfactory fits are obtained for T$_d$=24-19~K and dust spectral indexes $\beta$=1.4-1.8
respectively (see the best fit as continuous blue curve in Fig.~\ref{fig:sed}).
A luminosity of L$_{bol}$=5.5$\pm$0.5\,${L_\odot}$ and a ``bolometric temperature" of 
T$_{bol}$\,$\simeq$175\,K (the temperature of a blackbody with the same SED mean frequency; 
Myers \& Ladd 1993) are inferred from the full fit.

In a first inspection, the resulting SED  (Fig.~\ref{fig:sed}) does not look like a typical Class~0 
source in the near-- and mid--IR (see e.g., Whitney et al. 2003).  The improved sensitivity of the 
IR instrumentation  compared to that available in the 90's  has allowed the detection of  scattered 
light from several Class~0 sources below $\sim$10\,$\mu$m (Tobin et al. 2007).
In addition, the K$_s$--band emission due to scattered light observed 
through a cavity opened  by the outflow (with a favorable inclination angle)
and/or H$_2$ rovibrational emission from the outflow base, has also been
detected in a few Class~0 sources (e.g., Tachihara et al. 2007).
IRAS~22129+7000 shows these observable characteristics but it also displays an intriguing 
strong, and rather flat, near-- and mid--IR emission (as shown by the relatively high T$_{bol}$).
In fact, \textit{Spitzer} fluxes towards IRAS 22129+7000 are consistent
with  the typical values observed in Class~I sources with the same luminosity 
(Furlan et al. 2008), but an order of magnitude higher than in
Class~0 sources (Tobin et al. 2007). On the other hand,
IRAS 22129+7000 shows a very large fraction of
submm ($\lambda$\,$>$350\,$\mu$m) to bolometric luminosity ($\simeq$2$\%$; 
a signature of younger Class~0 sources still surrounded by massive dusty envelopes).

\clearpage

In an attempt to better understand the peculiar IR shape we tried to fit the SED with the 
2D radiative transfer tool by Robitaille et al. (2007). 
The best models fitting the data are shown in Fig.~\ref{fig:sed} for two different
inclination angles.  While the long-wavelength emission from the circumstellar envelope 
is well reproduced by the models (and is almost not affected
by the viewing angle), the  short--wavelength emission from the protostellar/disk 
system is clearly underestimated and strongly depends on inclination.
Therefore, a small viewing angle favors the detection of near-- and mid--IR scattered
light from the embedded  protostar through the cavities cleared by the outflow. 
However, the blue-- and red--shifted lobes of the CO outflow  (Fig.~\ref{fig:IRAC}) appear relatively 
well separated, and thus the source can not be completely ``pole--in'' \cite{Cab86}. 
In summary, the observed SED is peculiar, with 
a \textit{near-- and mid--IR excess} that can not be fully explained in terms of inclination effects. 

\subsection{Circumstellar Envelope: Radial Analysis}

\begin{figure}[hb]
\centering
\includegraphics[angle=-90,width=8.5cm]{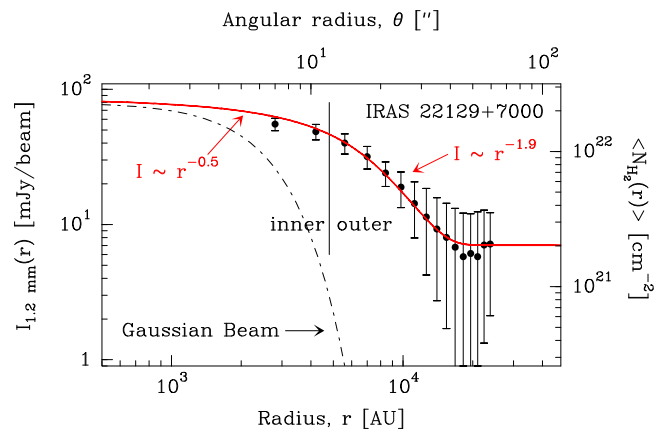}
\caption{1.2~mm radial intensity profile around IRAS~22129+7000.
The intensity (in mJy/11$''$-beam) has been averaged over circular annulus in steps of 3.5$''$.
The error bars show the standard deviation within each annulus.
The dashed line represent a Gaussian beam with a FWHM of 11$''$.
The right axis shows an estimate of the H$_2$ column density assuming
$\kappa_{1.2}$=0.01\,cm$^2$\,g$^{-1}$ and T$_d$=20~K. The red curve shows
a fit to the data including 2 intensity power-law profiles (convolved
with the beam), with $I(r)\propto r^{-1.9}$ for $r$\,$\gtrsim$5,000\,AU
and $I(r)\propto r^{-0.5}$ for $r$\,$\lesssim$5,000\,AU.}
\label{fig:profile}
\end{figure}

The cold dust surrounding the YSO has been mapped through 
its 1.2\,mm  emission (Fig.~\ref{fig:mambo}). 
As mm continuum emission is generally optically thin, 
the measured fluxes are proportional to the 
temperature--weighted mass content assuming that dust grains properties are known. 
 In  order to trace the mass distribution 
in the envelope, the radial intensity profile $I(r)$ has been analyzed by averaging the 
1.2\,mm fluxes within  equidistant circular  annulus (Fig.~\ref{fig:profile}). 
An envelope radius of $R_{out}$\,$\simeq$15,000\,AU is inferred. Assuming spherical symmetry, 
the intensity profile $I(r)\propto r^{-m}$ contains information on the density distribution, 
$\rho(r)\propto r^{-p}$, and temperature distribution, $T_d(r)\propto r^{-q}$, with $p=m+1-q$
(e.g., Adams 1991).

\begin{deluxetable}{ccccc}
\tabletypesize{\scriptsize}
\tablecaption{IRAS~22129+700: SED parameters
\label{tab-sed}}
\tablewidth{0pt}
\tablehead{ 
L$_{bol}$     & T$_{bol}$ & M$_{env}$         & L$_{smm}$/L$_{bol}$ & M$_{env}$/L$_{bol}$\\
(L$_{\odot}$) &   (K)     & (M$_{\odot}$)      &    (\%)$^c$     & (M$_{\odot}$/L$_{\odot}$) \\}
\startdata
5.5$\pm$0.5   & $\sim$175  & 0.2$^a$-0.7$^b$   &      $\sim$2       &  0.04$^a$-0.13$^b$\\
\enddata
\tablenotetext{a}{Assuming T$_d$=20\,K, $\kappa_{1.2}$=0.01\,cm$^2$\,g$^{-1}$ and $r\lesssim$5,000\,AU.}
\tablenotetext{b}{Mass within $r$=R$_{out}$\,$\lesssim$15,000\,AU.}
\tablenotetext{c}{L$_{smm}$ is the ``submm" luminosity radiated at $\lambda>$350~$\mu$m.}
\end{deluxetable}

Before convolving the expected intensity profile with the bolometer angular resolution, 
we divide the envelope in two regions (\textit{inner} and \textit{outer})
and try to fit the observed 1.2\,mm emission profile with two power-laws. 
Fig.~\ref{fig:profile} shows that the dust emission in the outer envelope 
($r$\,$\gtrsim$5,000\,AU $\gtrsim$\,one beam) follows a very steep profile, with $m$\,$\simeq$1.9,
 connecting the 
envelope with the lower density ambient cloud at $r\geq R_{out}$. 
Since the outer envelope angular radii are larger than the  beam's FWHM,
 the convolution does not modify the slope of the intensity profile 
(e.g., Adams 1991; Motte \& Andr\'e 2001). For the same reason, our observations 
can not resolve  accurately the slope of the inner envelope ($r$\,$<$\,$R_{in}$\,$\simeq$5,000\,AU).
We take $m$\,$\lesssim$0.5 as the expected value in a free-falling envelope.
The line radiative transfer models shown in Sect.~\ref{fig:RTmodel} are not inconsistent with this
assumption.

Taking again a dust opacity of $\kappa_{1200}$=0.01\,cm$^2$\,g$^{-1}$  
and  T$_d$=20\,K (consistent with the SED analysis)
we compute a total circumstellar mass of M$_{env}$=0.7\,${M_\odot}$ within $r\leq R_{out}$. 
However, it is unlikely that all this mass can be finally accreted by the protostar/disk.
First because the outflow will disperse a large fraction of it, and second, because differential
 rotation in the
envelope is expected to decouple the inner rapidly infalling and rotating layers
from the slowly collapsing and rotating outer regions (Belloche et al. 2002).
Hence, we also computed M$_{env}^{in}$=0.2\,${M_\odot}$, the mass within one 11$''$--beam 
($r$\,$\lesssim$\,R$_{in}$) as representative of the mass inside the inner envelope.

\clearpage

The steep slope of the 1.2\,mm intensity profile suggests that a 
temperature gradient is also present.
In particular, the  dust thermal profile through the inner envelope should be
dominated by accretion heating from the protostar/disk system
with \cite{Ter93}:
\begin{equation}
T_d(r) \approx 38\, \emr{K}\, \left(\frac{r}{100\emr{\,AU}} \right)^{-q}
          \, \left(\frac{\emr{L_{\star}}}{\emr{L_{\odot}}} \right)^{q/2}
\label{eq-Td}
\end{equation}
where $q$ depends on the dust grain properties through $q=2/(4+\beta)$.
Given the proximity of Ced~201 PDR, FUV photons from BD+69$^\circ $1231 star 
also heat the outer layers of IRAS~22129+7000 envelope.
Penetration of this external FUV radiation field into the envelope greatly depends on the 
scattering properties of the dust particles, with larger penetration depths as grains grow
toward bigger grains (Goicoechea \& Le Bourlot 2007).
In the following we adopt a temperature profile that follows Eq.~(\ref{eq-Td}), reaches a minimum of 
$\sim$10\,K at $r$\,$\sim$10,000\,AU and then increases up to $\simeq$15\,K at $R_{out}$
due to external FUV heating (this temperature is consistent with our analysis of CO lines in 
the ambient cloud). Taking $q$=0.36 (using $\beta$=1.6 from the SED)
and assuming $L_{\star}\simeq L_{bol}$, the estimated mass of the envelope increases by a factor
$\sim$2 respect to the isothermal case because $T_d(r)$  drops below 20\,K in a large fraction
of the envelope. This difference adds to the  dominant source of uncertainty in the mass
determination,  the dust composition and associated opacity.

Taking into account these caveats, \textit{the observed 1.2\,mm intensity profile in the 
outer envelope  is equivalent to a  steep density profile
  $\rho(r>R_{in}) \propto r^{-2.54}$} with a number density of  
$n$(H$_2$)$\simeq$(1-2)$\times$10$^5$\,cm$^{-3}$ at $R_{in}$.
Under these assumptions, the (beam averaged) extinction towards 
the peak (e.g., Motte \& Andr\'e 2001) is A$_V\simeq$20-30, 
a factor $\sim$10 larger than the value implied by our 1.2\,mm continuum
observations in the ambient cloud.


\subsection{Gas Kinematics: Outflow and Infall?}

Figure~\ref{fig:IRAC} shows the outflow emanating from IRAS~22129+7000
and extending $\sim$0.3\,pc in the NW direction (redshifted lobe) and 
$\sim$0.2\,pc in the SE direction (blueshifted lobe). 
In addition, IRAS~22129+7000 is the driving source of the 
Herbig Haro HH450 jet detected $\sim$0.2~pc SE from the object \cite{Bal01}. 
The CO outflow is well collimated, with an aspect ratio of $\sim$3--4 
(the projected FWHM flow length/width  ratio).
CO $J$=2--1 to 4--3 line profiles
show high velocity wing emission from accelerated gas.  
Typical full line widths  are consistent with an 
expansion velocity of $v_{\emr{exp}}\simeq$7\,km\,s$^{-1}$ (Fig.~\ref{fig:mambo}).
We thus estimate that the dynamical age of the outflow ($\sim$length/$v_{\emr{exp}}$) 
is $\sim$35,000\,yr (up to $\sim$10$^5$\,yr if the inclination angle  is $i$=18$^o$).
The presence of a molecular outflow suggests that the protostar/disk system
is still accreting material from the natal envelope.
The high density tracer HCO$^+$ $J$=1-0 line, with a critical density of
$n_{\emr{cr}}^{1-0}$$\simeq$$2\times10^5$\,cm$^{-3}$,
shows a deep red--shifted--absorption asymmetry in its optically thick line profile
(when $n(H_2)$\,$>$\,$n^{J-J'}_{\emr{cr}}$ the associated $J$ level is
predominantly populated by collisions with H$_2$ molecules and T$_{ex}$, the 
transition excitation temperature,  
tends to the kinetic temperature).
The depth and velocity position of the absorption dip are due to a decrease of T$_{ex}$
at the outer edges of a (likely) infalling envelope (\textit{left insets} in Fig.~\ref{fig:mambo}).
As low-$J$ CO lines easily thermalize at lower densities ($n_{\emr{cr}}$\,$\simeq$10$^{3-4}$\,cm$^{-3}$), 
they may not be reliable tracers of the presence of infall motions but of the varying 
physical conditions in the beam. Among the observed CO lines, only the $J$=4-3 line starts to trace
denser ($n_{\emr{cr}}^{4-3}$$\simeq$$5\times10^4$\,cm$^{-3}$),
possibly infalling, gas.
In anycase, the HCO$^+$ $J$=1-0 self-absorption dip is redshifted 0.3\,km\,s$^{-1}$ from the rest velocity
of the cloud ($-$5\,km\,s$^{-1}$) and is not observed at nearby positions of the
cloud  where the line shows a pure Gaussian emission  profile 
(\textit{right insets} in Fig.~\ref{fig:mambo}).
Although higher--$J$ optically thick lines from abundant high dipole moment molecules
should be mapped to draw definitive conclusions, in the following we assume that the observed 
deep HCO$^+$ self-absorption is produced by inward motions.  

The densest gas in the envelope has been traced with the
DCO$^+$ $J$=3--2 line ($n_{\emr{cr}}^{3-2}$$\simeq$$2\times10^6$\,cm$^{-3}$) which shows 
compact ($\lesssim$40$''$) emission around IRAS~22129+7000. 
Compared to the broad line wings from the CO outflow, the DCO$^+$ $J$=3-2 
line shows the narrowest observed linewidth ($\Delta v_{\emr{FHWM}}$=0.60$\pm$0.04\,km\,s$^{-1}$).
On the other hand, we used the  C$^{18}$O $J$=1--0 line to trace lower density gas 
(e.g., the outer envelope). This line displays a broader linewidth 
($\Delta v_{\emr{FHWM}}$=0.96$\pm$0.02\,km\,s$^{-1}$).
The radiative transfer calculations presented in the next section confirm that both lines are 
optically thin and thus they are not broaden by line saturation. Therefore, the different
velocity dispersions reflect the distinct regions where both line profiles are formed.
Assuming  T$_k$=10 to 20\,K, the observed DCO$^+$ linewidths are 5 to 3.5 times larger than
the expected thermal broadening. The required non--thermal
velocity dispersion  to fit the observed DCO$^+$ $J$=3-2 linewidth (from turbulence, infall or rotation)
is $\sigma_{nth}\simeq$0.25\,km\,s$^{-1}$ (with $\Delta v_{\emr{FHWM}}=2.355\times\sigma$). 

\subsection{Line Radiative Transfer Analysis}
\label{fig:RTmodel}

In order to reproduce the observed molecular  lines with a realistic
model of the envelope,  we have used a nonlocal and non-LTE 
excitation and radiative transfer code \cite{Goi06}.
The code models an infalling spherical envelope and accounts for line trapping,
collisional excitation\footnote{CO (and isotopologues) rotationally inelastic collisional
rates have been derived from those of Flower (2001).
HCO$^+$ and DCO$^+$ collisional rates have been derived from those
of Flower (1999).}, and radiative excitation.
Arbitrary velocity fields as well as density, temperature, and abundance radial gradients
are included as input parameters. This is done by numerical discretization of the 
envelope in spherical shells.
The nonlocal radiative transfer problem is then simulated by the emission of a determined number
of model photons (cosmic mm background, continuum and line photons) using the
Monte Carlo approach \cite{Ber79}. The steady state statistical equilibrium equations
are then solved iteratively and the non-LTE level populations are determined in the shells.
The emergent line intensities along each line of sight are finally convolved with the
telescope angular resolution.

\begin{figure*}[ht]
\centering
\includegraphics[angle=-90,width=14cm]{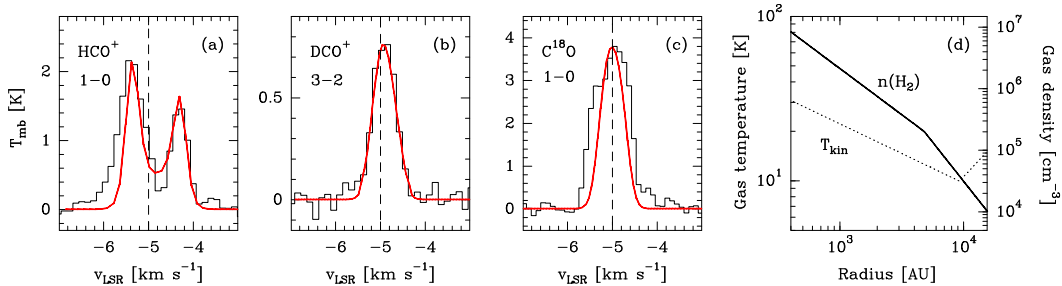}
\caption{$(a)$ HCO$^+$ $J$=1--0, $(b)$ DCO$^+$ $J$=3--2 and $(c)$ C$^{18}$O $J$=1--0 
observations (histograms) towards IRAS~22129+7000. Best radiative transfer models
for the circumstellar envelope are also shown (red curves). Predicted line
profiles have been convolved with the telescope angular resolution at each frequency.
Intensity scale is in main beam temperature.  
$(d)$ Density and temperature profiles used in the model.}
\label{fig:mtc}
\end{figure*}

\subsubsection{A First Model for the Circumstellar Envelope}

As our first attempt to understand  IRAS~22129+7000 physical structure, we have
modelled a spherical envelope with a radius $R_{out}$=15,000\,AU. We take
 the density and temperature profiles ($\rho(r)\propto r^{-2.54}$  
and $T(r)\propto r^{-0.34}$ plus external FUV heating) derived in the previous 
section for the outer envelope ($r$\,$>$\,$R_{in}$\,$\simeq$5,000\,AU). 
Given the high densities expected in most of the envelope, we
assume that gas and dust are thermalized (i.e., $T_k$=$T_d$).

Using the above density and temperature structure  we run the model globally for 
HCO$^+$, DCO$^+$ and C$^{18}$O and try to find iteratively the best velocity fields and abundances
fitting the observed line profiles.
In order to reproduce the observed velocity of the CO $J$=4-3 and 
HCO$^+$ $J$=1-0 redshifted absorption dips (0.2 and 0.3\,km\,s$^{-1}$) we adopt a uniform
infalling velocity of $v_{\emr{inf}}$=0.2\,km\,s$^{-1}$ in the outer envelope.
For $r\le R_{in}$ (the inner envelope) we assume $\rho(r)\propto r^{-1.5}$,
and $v_{\emr{inf}}(r)=0.2\, (r/R_{in})^{-0.5}$ i.e., the density and velocity
profiles in a freefall inside-out collapsing envelope \cite{Shu87}.
Numerical simulations of collapse induced by external compression waves 
also predict those approximate $\rho(r)$ and $v_{\emr{inf}}(r)$ profiles 
behind the expanding collapse-wavefront (e.g., Hennebelle et al. 2003).
These two regimes of infall (freefall and uniform) have been inferred in other YSOs with clear
evidences of collapse (Belloche et al. 2002). 
Under the above conditions, DCO$^+$ optically thin lines
constrain the remaining nonthermal broadening 
(\textit{turbulence} in the following).
Satisfactory fits to the DCO$^+$ $J$=3-2 line   (with [DCO$^+$/H$_2$]=(1-2)$\times$10$^{-10}$)
are obtained for $\sigma_{turb}$=0.15\,km\,s$^{-1}$ (the turbulence on the size scale of the beam).
Therefore, $\sigma_{turb}$ in the densest parts of the envelope (as traced by the DCO$^+$) is subsonic,
and  probably a vestige of the low degree of turbulence found in prestellar cores \cite{Berg07}.
Hereafter we adopt a uniform $\sigma_{turb}$ value of 0.15\,km\,s$^{-1}$
in the models of the envelope.

In order to reproduce the HCO$^+$ $J$=1--0 line profile we have
added the contribution from the lower density (more turbulent) ambient cloud 
surrounding IRAS~22129+7000  and producing a line-of-sight extinction of A$_V$=2-3 
(from 1.2\,mm observations). 
Given the high HCO$^+$ dipole moment and large abundance 
(producing subthermal excitation and large opacities),  low-$J$ line photons from the dense  
envelope will be greatly attenuated 
and scattered over large areas by this low density \textit{halo} \cite{Cer87}.
Note that the negligible DCO$^+$ and  C$^{18}$O column densities in the ambient cloud
do not alter the DCO$^+$ and  C$^{18}$O line emission from the envelope itself.
The density and temperature in the ambient cloud can be constrained with the
observed CO lines. They are reproduced with   
n(H$_2$)$\simeq$1000\,cm$^{-3}$ and T$_k$\,$\simeq$15\,K, 
in agreement with estimations by Witt et al. (1987).
 The \textit{halo} ambient cloud in the model is at rest 
($v_{\emr{inf}}$=0\,km\,s$^{-1}$) but the broader observed linewidths 
require $\sigma_{turb}$=0.30\,km\,s$^{-1}$ (supersonic), at least a factor 2 larger
than in the circumstellar envelope.
Figure \ref{fig:mtc} shows the best fits and the density and temperaure gradients used
in the model. In spite of the good agreement, the model slightly underestimates the observed  
HCO$^+$ and C$^{18}$O $J$=1--0 linewidths, likely due to a minor contribution from the 
high velocity outflow (not modelled here).

The resulting large DCO$^+$/HCO$^+$ abundance ratio ($\sim$0.02) found in
 IRAS~22129+7000 
is similar to that derived in cold and dense prestellar condensations near other PDRs 
(e.g., Pety et al. 2007). This is consistent for a species that shows enhanced abundance 
in the cold FUV--shielded gas (low ionization degree) affected  by ongoing deuterium fractionation 
(where dissociative recombinations with electrons are its main destruction route).
As a consequence, the DCO$^+$/HCO$^+$ abundance ratio is roughly inversely proportional
to the electrons abundance and we can estimate its value.
Taking the same gas phase chemical network used by us in Pety et al. (2007), 
and assuming a cosmic ionization rate of 
$\zeta$=3$\times$10$^{-17}$\,s$^{-1}$ \cite{Dal06},  we derive an ionization fraction
of $\sim$10$^{-8}$ (with respect to H nuclei) in the densest and coldest 
regions of the envelope where the DCO$^+$ abundance peaks.
This value is, within a factor $\sim$2, similar 
to the ionization fraction
inferred in prestellar cores such as Barnard~68 \cite{Mar07}.
 
\subsubsection{CO outflow parameters}

In the previous section we have estimated the amount of material towards IRAS~22129+7000
and also evaluated the CO line emission contribution from the ambient cloud 
and from the circumstellar envelope (through the optically thin C$^{18}$O line). 
After subtracting the CO contribution from the above  line--of--sight components we now estimate a 
CO column density of $\sim$10$^{16}$\,cm$^{-2}$ in the outflow.  
Taking into account the observed dimensions of the blue-- and red--shifted
lobes we  compute the mass ($\sim$0.14\,M$_{\odot}$), momentum ($\sim$1.0\,M$_{\odot}$\,km\,s$^{-1}$)
and kinetic energy ($\sim$6.8$\times$10$^{43}$\,erg)
carried out by the outflow.
Note that the mechanical luminosity  of the flow ($\sim$0.02\,L$_{\odot}$) is
still a significant fraction ($\sim$0.4\%) of the radiant luminosity 
of the driving source. The estimated outflow momentum flux is
F$_{\emr{CO}}$\,$\sim$3$\times$10$^{-5}$\,M$_{\odot}$\,km\,s$^{-1}$\,yr$^{-1}$,
although the  correction factor for inclination effects ($sin\,i /cos^2\,i$) will decrease 
the actual F$_{\emr{CO}}$ if $i<$40$^o$. 
Hence, F$_{\emr{CO}}$ is intermediate between the observed momentum fluxes found
in early  Class~0 sources and those found in late Class~I sources  
(e.g,  Bontemps et al. 1996, Bachiller 1996).

\begin{deluxetable}{lcccc}[hb]
\tabletypesize{\scriptsize}
\tablecaption{IRAS~22129+700: CO outflow parameters
\label{tab-outflow}}
\tablewidth{0pt}
\tablehead{ 
              &               &                             & Kinetic          & Mechanical\\
Outflow       &      Mass     &         Momentum            & Energy           & Luminosity\\
Component$^a$ & (M$_{\odot}$) & (M$_{\odot}$\,km\,s$^{-1}$) & (10$^{43}$\,erg) & (10$^{-3}$\,L$_{\odot}$)\\}
\startdata
Blueshifted...&  0.06         &          0.4                &   2.9            &  6.9      \\
Redshifted....&  0.08         &          0.6                &   3.9            &  9.3      \\
Total............&  0.14      &          1.0                &   6.8            &  16.2     \\
\enddata
\tablenotetext{a}{To calculate the energy and momentum we use an expansion velocity  of 7\,km\,s$^{-1}$
and a dynamical age of 35,000~yr (see text).}
\end{deluxetable}

\section{Discussion and Conclusions}

\subsection{The Nature of IRAS~22129+7000}
As noted earlier IRAS~22129+7000 shows indirect evidence of a
protostar/disk system (a collimated outflow) and it shows  extended dust emission
from the envelope with a large L$_{smm}$/L$_{bol}$ ratio ($\simeq$2$\%$).
These characteristics are consistent with the standard observational definition
 of Class~0 sources (Andr\'e et al. 2000). Comparison of the derived L$_{bol}$ and 
M$_{env}^{in}$ values  with theoretical 
evolutionary tracks \cite{Sar96,And00} shows that
the source is likely in the interesting Class 0 to I transition. Approximately
50$\%$ of the initial mass envelope must have been accreted at this stage. 
According to these diagrams, infall must have started in the parental core 
$\sim$(0.5-1)$\times$10$^5$~yr ago. This age is consistent with the dynamical age of
the outflow, specially if inclination effects are taken into account.

The peculiar SED of IRAS~22129+7000 shows however a \textit{near-- and mid--IR excess} that produces a high T$_{bol}$
of 175\,K, consistent with more evolved Class~I sources ($\sim$70--650\,K; Myers et al. 1998).
Given the strong dependence of the short-wavelength IR emission on the viewing angle 
(Fig.\ref{fig:sed}), the inferred T$_{bol}$  is 
certainly affected by orientation effects. Up-to-date 2D SED models
show however that inclination is not enough to reproduce the observed \textit{excess}.
Nevertheless, as mentioned by Robitaille et al. (2006), their large grid of modelled SEDs
have some caveats. The most important for our analysis is that none of them 
includes multiple central sources.
An interesting possibility is thus that the source in IRAS~22129+7000 is a multiple 
(e.g., binary) stellar system.
In fact, prestellar cores can fragment a second time during protostellar collapse and lead
to the formation of binary systems (e.g. Duch\^ene et al. 2007). 
However, it is not yet clear if the initial conditions in these cores (e.g., the cloud environment) 
play a major role in such  dynamical fragmentation. 
Simulations predict that external compression waves 
arriving at the central object at different times 
or instabilities in the outer accretion disk could be the seed of multiple protostellar systems
(e.g., Hennebelle et al. 2003, 2004).
In this picture, the presence
of both a young and a more evolved YSO sharing a common circumstellar envelope 
would explain the observed \textit{near-- and mid--IR excess} if the scattered light from the
evolved YSO escapes from the outflow cavities.
Higher angular resolution IR observations are  required to confirm  this scenario.
In addition, a larger set of molecular line observations and maps are needed 
to tightly constrain the physics (infall and rotation velocity fields) and 
chemistry (molecular content) of this protostar and its environment.

\subsection{Star Formation Near Ced~201 PDR}

IRAS~22129+7000 is located near  Ced~201 PDR ($d$\,$\sim$0.4\,pc and A$_V$\,$\sim$1).
Taking into account geometrical dilution and  selective dust extinction
of FUV dissociating radiation from BD+69$^\circ $1231, Casey (1991) computed
that a region with a radius of at least $\sim$0.6\,pc ($\sim$300$''$) around the 
exciting star is permeated by FUV photons.
In particular the FUV radiation field near IRAS~22129+7000 should be presently at least
$\sim$2 times the mean interstellar radiation field.
Aside from dynamical effects  associated with the presence of a nearby PDR
illuminated by a runaway star (e.g., shock waves), heating and radiation pressure in the region are 
larger than in other FUV radiation--free environments. 
Interestingly enough, the 1.2\,mm dust emission (very weak in the PDR itself) peaks
again in two dust condensations $\sim$0.25\,pc North of Ced~201, roughly in the
same direction of  outflow red--lobe axis.
These cores/clumps have a radius of $\sim$5,000\,AU and coincide with an enhancement of the 
HCO$^+$ $J$=1--0 line emission (see Fig.\ref{fig:mambo}).
It is very likely that  these condensations 
are being externally heated and compressed by  the combined action of the outflow and the PDR
associated  FUV--driven shock wave. In fact, a similar scenario has presumably led to the
formation of  IRAS~22129+7000 itself. We thus suspect that star formation in the B175 globule 
is triggered by the shocks and radiative feedback induced 
by BD+69$^\circ $1231 star and the new generation of low-mass YSOs.

Observations of cores and YSOs provide new insights into how the star formation process
is affected by its cloud environment. The vicinity of late type B stars
has interesting intermediate properties between isolated dark cores (where FUV-radiation does not 
play a major role) and  cores compressed by the expansion of  H\,{\sc ii} regions 
(near more massive stars). Cool PDRs such as Ced~201 are 
nearly devoid of ionizing photons and thus they provide a very different environment 
compared to other star forming regions which are photoionized by brighter OB stars. 
In the special case of Ced~201, the exciting star is moving fast
through the region, suggesting that its arrival to the cloud, and
thus the triggering,  is recent.

\acknowledgments
We thank P. Hennebelle for useful advice, J. Pety for  his help with CLASS90 and the 
SPECPDR team for their contribution to the project. We also thank the referee
for his/her constructive  criticisms.
JRG was supported by  a \textit{Marie Curie intra-European Individual Fellowship},
contract MEIF--CT--2005--515340.
We made use of data products from the 2MASS Catalog, 
which is a joint project of the U. of Massachusetts and 
the IPAC/Caltech, funded by NASA and the NSF."


\begin{thebibliography}{}


\bibitem[Adams 1991]{Ada91}
Adams, F. C. 1991, ApJ, 382, 544

\bibitem[Andr\'e et al. 1993]{And93}
Andr\'e, P., Ward-Thompson, D. \& Barsony, M. 1993,  ApJ, 406, 122

\bibitem[Andr\'e et al. 2000]{And00}
Andr\'e, P., Ward-Thompson, D. \& Barsony, M. 2000,
  Protostars and Planets IV, 
University of Arizona Press, eds Mannings, V., Boss, A.P., Russell, S. S., p59

\bibitem[Arce \& Sargent 2006]{Arc06}
Arce, H,G. \& Sargent, A.I. 2006, ApJ, 646, 1070 

\bibitem[Bally \& Reipurth 2001]{Bal01}
Bally, J. \&  Reipurth, B. 2001, ApJ, 552, L159

\bibitem[Bachiller 1996]{Bac96}
Bachiller, R. 1996, ARA\&A, Volume 34, 111-154

\bibitem[Belloche et al. 2002]{Bel02}
Belloche, A., Andr\'e, P.,  Despois, D. \& Blinder, S. 
2002, A\&A, 393, 927

\bibitem[Bergin \& Tafalla 2007]{Berg07}
Bergin, E.A., \&  Tafalla, Mario, 2007, ARA\&A, 45, 339

\bibitem[Bern\'e et al. 2007]{Ber07}
Bern\'e, O. et al. 2007, A\&A, 469, 575

\bibitem[Bernes 1979]{Ber79}
Bernes, C. 1979, A\&A, 73, 67.


\bibitem[Bontemps et al. 1996]{Bon96}
Bontemps, S., Andr\'e, P., Terebey, S. \& Cabrit, S.
1996, A\&A, 311, 858

\bibitem[Brinch et al. 2007]{Bri07}
Brinch, C., Crapsi, A., Hogerheijde, M. R. \& J$\o$rgensen, J. K. 2007, A\&A, 461, 1037

\bibitem[Cabrit \& Bertout 1986]{Cab86}
Cabrit, S. \& Bertout, C.
1986,  ApJ, 307, 313


\bibitem[Cabrit et al. 2007]{Cab07}
Cabrit, S., Codella, C., Gueth, F., Nisini, B., Gusdorf, A., Dougados, C. \& Bacciotti, F.
2007,  A\&A, 468, L2

\bibitem[Casey 1991]{Cas91}
Casey, S.C. 1991, ApJ, 371, 183

\bibitem[Cernicharo \& Gu\'elin 1987]{Cer87}
Cernicharo, J. \& Gu\'elin, M. 1987, A\&A, 176, 299


\bibitem[Cesarsky et al. 2000]{Ces00}
Cesarsky, D., Lequeux, J., Ryter, C., Gerin, M. 2000, A\&A, 354, L87


\bibitem[Dalgarno 2006]{Dal06}
Dalgarno, A., 2006 \textit{Proceedings of the National Academy of Science}, vol. 103, Issue 33, 12269-12273


\bibitem[Duch\^ene et al. 2007]{Duc07}
Duch\^ene, G., Bontemps, S., Bouvier, J., Andr\'e, P., Djupvik, A. A. \& Ghez, A. M.
2007, A\&A, 476, 229


\bibitem[Flower 1999]{Flo99}
Flower, D. R.
1999, MNRAS, 305, 


\bibitem[Flower 2001]{Flo01}
Flower, D. R.
2001, JPhB, 305, 651


\bibitem[Furlan et al. 2008]{Fur08}
Furlan E. et al., 2008, to appear in ApJS, astro-ph/07114038. 


\bibitem[Goicoechea \& Le Bourlot 2007]{Goi07}
Goicoechea, J. R. \& Le Bourlot, J. 2007, A\&A, 467, 1


\bibitem[Goicoechea et al. 2006]{Goi06}
Goicoechea, J. R., Pety, J., Gerin, M., Teyssier, D., Roueff, E., Hily-Blant, P. \& Baek, S.
2006, A\&A, 456, 565

\bibitem[Hennebelle et al. 2003]{Hen03}
Hennebelle, P., Whitworth, A. P., Gladwin, P. P., \& Andr\'e, Ph.
2003, MNRAS, 340, 870


\bibitem[Hennebelle et al. 2004]{Hen04}
Hennebelle, P., Whitworth, A. P., Cha, S.-H., \& Goodwin, S. P. 
2004, MNRAS, 348, 687.

\bibitem[Joblin et al. 2005]{Job05}
Joblin, C. et al. 2005, on \textit{Astrochemistry: Recent Successes and Current Challenges},
Proceedings of the 231st Symposium of the IAU held in Pacific Grove, California, USA, p.194


\bibitem[Johnstone et al. 2000]{Joh00}
Johnstone, D., Wilson, Ch.D., Moriarty-Schieven, G., Joncas, G.,
Smith, G., Gregersen, E. \& Fich, M. 2000, ApJ, 545, 327

\bibitem[J$\o$rgensen et al. 2007]{Jor07}
J$\o$rgensen, J.K. et al. 2007, ApJ, 659, 479

\bibitem[Kun 1998]{Kun98}
Kun, M. 1998, ApJS, 115, 59

\bibitem[Lada 1999]{Lad99}
Lada, C.J. 1999, 
The Origin of Stars and Planetary Systems. 
Edited by Charles J. Lada and Nikolaos D. Kylafis. 
Kluwer Academic Publishers, 1999, p143

\bibitem[Maret \& Bergin 2007]{Mar07}
Maret, S, \&  Bergin, E.A. 
2007, ApJ, 664, 956



\bibitem[Motte \& Andr\'e]{Mot01}
Motte, F. \&  Andr\'e, P. 2001, A\&A, 440, 464




\bibitem[Myers \& Ladd 1993]{Mye93}
Myers, P. C. \& Ladd, E. F. 1993, 413, L4

\bibitem[Myers et al. 1998]{Mye98}
Myers, P. C., Adams, F. C., Chen, H. \& Schaff, E. 1998, ApJ, 492, 703


\bibitem[Nikoli\'c \& Kun 2004]{Nik04}
Nikoli\'c, S. \& Kun, M. 2004, BaltA, 13, 487

\bibitem[Ossenkopf \& Henning 1994]{Oss94}
Ossenkopf, V. \& Henning, Th. A\&A, 291, 943


\bibitem[Pety et al. 2007]{Pet07}
Pety, J., Goicoechea, J. R., Hily-Blant, P., Gerin, M., Teyssier, D. 2007, A\&A, 464, L41


\bibitem[Robitaille et al. 2006]{Rob06}
Robitaille, T.P., Whitney, B.A., Indebetouw, R., Wood, K. \& Denzmore, P. 2006, ApJS, 167, 256

\bibitem[Saraceno et al. 1996]{Sar96}
Saraceno, P., Andr\'e, P., Ceccarelli, C., Griffin, M. \& Molinari, S.
1996, A\&A, 309, 827

\bibitem[Shu et al. 1987]{Shu87}
Shu, F,H., Adams, F.C. \& Lizano, S. 
1987, ARA\&A, 25, 23


\bibitem[Skrutskie et al. 2006]{Skr06}
Skrutskie et al. 2006, AJ, 131, 1163

\bibitem[Smith et al. 2007]{Smi07}
Smith, J. D. T. et al. 2007, PASP, 119, 1133 

\bibitem[Tachihara et al. 2007]{Tac07}
Tachihara, K. et al. 2007, ApJ, 659, 1382

\bibitem[Terebey et al. 1993]{Ter93}
Terebey, S., Chandler, C. J. \&  Andr\'e, P. 
1993, ApJ, 414, 759


\bibitem[Teyssier \& Sievers 1999]{Tey99}
Teyssier, D. \& Sievers, A. 1999. A Fast--Mapping Method for Bolometer
Arrary Observations. IRAM technical report.

\bibitem[Tobin et al. 2007]{Tob07}
Tobin, J.J., Looney, L.W., Mundy, L.G., Kwon, W., Hamidouche, M.  2007, ApJ, 659, 1404



\bibitem[Whitney et al. 2003]{Whi03}
Whitney, B,A., Wood, K., Bjorkman, J. E., Cohen, M. 2003, ApJ, 598, 1079


\bibitem[Witt et al. 1987]{Wit87}
Witt, A. N., Graff, S. M., Bohlin, R. C. \& Stecher, T. P. 1987, ApJ, 321, 912

\bibitem[Wolf 1908]{Wol08}
Wolf, M. 1908, MNRAS, 69, 117


\bibitem[Zylka 1998]{Zyl98}
Zylka, R. 1998, Pocket Cookbook for MOPSIC Software.


\end{thebibliography}
\end{document}